\documentclass[journal]{IEEEtran}
\usepackage{amsmath, amssymb, bbm}
\usepackage{bm}
\usepackage{amsthm}
\usepackage{graphicx,epsfig}
\usepackage{subfigure}
\usepackage{verbatim}
\usepackage{soul}
\usepackage[colorlinks]{hyperref}
\usepackage{color}
\usepackage{booktabs}
\usepackage{makecell}
\usepackage{graphicx}
\usepackage{enumitem}

\usepackage{algorithm}
\usepackage{algpseudocode}

\newtheorem*{proof*}{Proof}

\usepackage{hhline}
\usepackage{multirow}
\usepackage{caption}

\usepackage{cite}

\usepackage[utf8]{inputenc}
\usepackage{amssymb}
\usepackage{nomencl}
\usepackage{etoolbox}

\begin{document}
\title{Gaussian Process-based Approach for Bilevel Optimization in the power system - A Critical Load Restoration Case}
\author{Yang Liu$^{1}$,\textit{ Member}, \textit{IEEE},  Yu Weng$^{1}$,~\IEEEmembership{Student~Member,~IEEE,} Rufan Yang$^{1}$,~\IEEEmembership{Student~Member,~IEEE,} Quoc-Tuan Tran$^{2}$,~\IEEEmembership{Senior~Member,~IEEE,} and Hung D. Nguyen$^{1,\star}$, \textit{Member}, \textit{IEEE}

\thanks{ Yang Liu and Hung D. Nguyen, e-mail: \{liu-yang, hunghtd\}@ntu.edu.sg.

\textsuperscript{\,} Corresponding author: Hung D. Nguyen}}

\markboth{IEEE Transactions on XX,~Vol.~, No.~, }{}%

\maketitle

\begin{abstract}
Bilevel optimization problems can be used to represent the collaborative interaction between a power system and grid-connected entities, called the followers, such as data centers. Most existing approaches assume that such followers' response behaviors are made available to the power system in the operation decision-making, which may be untenable in reality. This work presents a novel idea of solving bilevel optimization problems without assuming the omniscience of the power system. The followers' responses will be represented by a function of the power system decisions using Gaussian Process Regression. Then the two layers in the bilevel problem can be solved separately by the power system and its followers. This not only avoids the omniscience assumption, but also significantly increases the computational efficiency without compromising accuracy, especially for the problems with a complex lower layer. Moreover, a bilevel critical load restoration model is developed to test the proposed technique. Compared to the conventional methods, the proposed restoration model considers the load-side operation and the varying load marginal value, and can accurately estimate load-side loss and achieve better restoration solutions. Two case studies validate the advantages of the proposed approaches from different perspectives. 
\end{abstract}


\makenomenclature
\renewcommand\nomgroup[1]{%
  \item[\bfseries
  \ifstrequal{#1}{A}{Indices}{%
  \ifstrequal{#1}{B}{Sets}{%
  \ifstrequal{#1}{C}{Parameters}{%
    \ifstrequal{#1}{D}{Variables}{%
  \ifstrequal{#1}{E}{Functions}{}}}}}%
]}

\nomenclature[A,4]{$t,t'$}{Index of periods}
\nomenclature[A,2]{$i,j$}{Index of nodes}
\nomenclature[A,3]{$idc$}{Index of IDCs}
\nomenclature[A,5]{$wl$}{Index of workloads in IDCs}
\nomenclature[B,1]{$\mathcal{B}$}{Set of all nodes}
\nomenclature[B,2]{$\mathcal{B}^{cl}$}{Set of critical nodes}
\nomenclature[B,3]{$\mathcal{B}^{idc}$}{Set of IDC nodes}
\nomenclature[B,4]{$\mathcal{IDC}$}{Set of IDCs located in a node}
\nomenclature[B,5]{$\mathcal{L}$}{Set of lines between two nodes}
\nomenclature[B,6]{$\mathcal{L^d}$}{Set of damaged lines between two nodes}
\nomenclature[C,11]{\textit{Upper Layer}}{}
\nomenclature[C,15]{$N$}{Number of nodes}
\nomenclature[C,16]{$N^R$}{Number of repairing crews}
\nomenclature[C,22]{$T^{r}_{i,j}$}{Required periods for repairing the line between node $i$ and $j$}
\nomenclature[C,17]{$P^d_{i,t}$}{Real power demand of node $i$ at time $t$}
\nomenclature[C,19]{$Q^d_{i,t}$}{Reactive power demand of node $i$ at time $t$}
\nomenclature[C,18]{$\underline{P^g_i}$,$\overline{P^g_i}$}{Minimum and maximum real power generation of node $i$}
\nomenclature[C,19]{$\underline{Q^g_i}$,$\overline{Q^g_i}$}{Minimum and maximum reactive power generation of node $i$}
\nomenclature[C,24]{$\underline{V_i}$,$\overline{V_i}$}{Voltage lower and upper limits of node $i$}
\nomenclature[C,20]{$r_{i,j}$}{Resistance of the line between node $i$ and $j$}
\nomenclature[C,24.5]{$x_{i,j}$}{Reactance of the line between node $i$ and $j$}
\nomenclature[C,21]{$S_{i,j}^{max}$}{Power Flow limit of the line between node $i$ and $j$}
\nomenclature[C,23]{$u^{sw}_{i,j,t}$}{Switch indicator of the line between node $i$ and $j$ at time $t$ (binary, 1 = equiped with remotely controlled switch, 0 = otherwise)}
\nomenclature[C,12]{$a_i,b_i,c_i$}{Coefficients in cost functions of node $i$}
\nomenclature[C,14]{$M$}{A very large number}
\nomenclature[C,13]{$m$}{A very small number}

\nomenclature[C,25]{\textit{Lower Layer}}{}
\nomenclature[C,33]{$W_{wl}$}{Worth of workload $wl$}
\nomenclature[C,28]{$RT_{wl}$}{Release time of workload $wl$}
\nomenclature[C,26]{$DL_{wl}$}{Deadline of workload $wl$}
\nomenclature[C,32]{$v^l_{wl,idc}$}{Location indicator of workload $wl$ (binary, 1 = located in IDC $idc$, 0 = otherwise)}
\nomenclature[C,30]{$S^{req}_{wl}$}{Required server resource amount of workload $wl$}
\nomenclature[C,29]{$S^{cap}_{idc}$}{Server resource capacity of IDC $idc$}
\nomenclature[C,27]{$LF_{idc}$}{Load factor of IDC $idc$}
\nomenclature[C,31]{$UP^{max}_{idc}$}{Maximum amount of uploaded workloads from IDC $idc$}
\nomenclature[D,11]{\textit{Upper Layer}}{}
\nomenclature[D,17.2]{$\rho^{s}_{i,t}$}{Load shedding percentage of node $i$ at time $t$}
\nomenclature[D,13]{$P^g_{i,t}$}{Real power generation of node $i$ at time $t$}
\nomenclature[D,15]{$Q^g_{i,t}$}{Reactive power generation of node $i$ at time $t$}
\nomenclature[D,12]{$P_{i,j,t}$}{Real power flow of the line between node $i$ and $j$ at time $t$}
\nomenclature[D,14]{$Q_{i,j,t}$}{Reactive power flow of the line between node $i$ and $j$ at time $t$}
\nomenclature[D,17.1]{$V_{i,t}$}{Voltage magnitude of node $i$ at time $t$}
\nomenclature[D,16]{$u_{i,j,t}$}{Connection status indicator of the line between node $i$ and $j$ at time $t$ (binary, 1 = connected, 0 = otherwise)}
\nomenclature[D,17]{$u^r_{i,j,t}$}{Reparation indicator of the line between node $i$ and $j$ at time $t$ (binary 1 = under reparation, 0 = otherwise)}
\nomenclature[D,18]{\textit{Lower Layer}}{}
\nomenclature[D,22]{$v^t_{wl}$}{Termination indicator of workload $wl$ (binary, 1 = terminated, 0 = otherwise) }
\nomenclature[D,21]{$v^c_{wl,idc}$}{Completion indicator of workload $wl$ (binary, 1 = completed in IDC $idc$, 0 = otherwise) }
\nomenclature[D,23]{$v^u_{wl,idc,t}$}{Upload indicator of workload $wl$ (binary, 1 = uploaded to cloud from IDC $idc$ at time $t$, 0 = otherwise) }
\nomenclature[D,20]{$S_{wl,idc,t}$}{Server resource allocated to workload $wl$ in IDC $idc$ at time $t$}
\nomenclature[D,19]{$L^{idc}$}{Total loss of IDCs during restoration periods}
\nomenclature[E]{$C_i$}{Cost function describing the load-side loss of node $i$}
\textcolor{black}{\printnomenclature}

\section{Introduction} 

Bilevel programming is widely used to represent the gaming problems in the power system, such as the service pricing problems in deregulated electricity markets \cite{bilevel_tariff}\cite{bilevel_demandresponse} and the collaborative optimization problem between the power system and natural gas systems \cite{bilevel_gas1}\cite{bilevel_gas2}. By modeling the operation of different parties in separate layers through bilevel programming, the interdependence between the parties is described by the interaction of the two layers, so that the operation strategies of each party considering the interdependence can be obtained by solving the bilevel model.

Stackelberg bilevel model is the most common bilevel gaming problems that the leader(s) moves first and then the follower(s) moves accordingly \cite{stackelberg_first,game_theory}. In power system operation, the power system is usually the game leader, whose decisions will affect the followers as grid-connected entities. By solving Stackelberg bilevel models, the optimal decisions of the power system considering the follower's response can be found. Apparently, one of the major assumptions made in such cases is that the power system are aware of all possible responses of the followers. In fact, the omniscience of the leader is the fundamental assumption for Stackelberg games \cite{bilevel_review} and most of existing solution methods of Stackelberg games \cite{single_level1,single_level2,descent1,descent2,trust_region1,trust_region2}.

However, the assumption of leaders' awareness to followers' response is not always tenable in the power system. In other words, there is a lack of information exchange between the layers due to, for example, privacy concerns and unobservability. For some cases such as the gaming between ISO and generators, ISO may own the knowledge of generators' bidding strategies through the analysis of their cost and bidding history. But for the cases that the followers are not managed or supervised by the power system such as natural gas systems and city heating systems, the power system may not acquire sufficient information required for the bilevel model, especially when some of the information such as natural gas network details are confidential \cite{gas_confidential}. For such cases, the existing approaches based on the leader's omniscience assumption are not directly applicable. Not to mention that the bilevel optimization may be a NP-hard problem \cite{bilevel_review} that incur a higher computational cost to solve. Therefore, developing an efficient approach that is not based on the assumption of the leader's omniscience becomes more necessary.

This paper introduces a novel method for solving bilevel collaborative optimization with Stackelberg gaming forms arising in the power system. Unlike the existing solution methods for Stackelberg bilevel problems, the proposed method does not assume the omniscience of the leader, meaning the follower's response modeled in the lower layer of the bilevel model is not aware by the leader. Instead, the follower's optimal response will be explicitly expressed as a function of the leader's decisions. This function will be provided by the followers to the leader, and then the upper and lower layer models can be solved separately. The major challenge in the proposed method is how to establish such function between the follower's optimal responses and the leader's decisions. Thus, Gaussian Process Regression (GPR) is leveraged to approximate how the follower respond to the leader's decision. Another contribution of the paper is a new critical load (CL) restoration bilevel scheme that considers the load-side operation and the varying marginal load value (MLV). This new scheme helps achieve more realistic solutions compared to the conventional methods. The bilevel restoration model will be used to demonstrate and test the proposed GPR-based solution method. The simulation results of two case studies not only show the advancement of the CL restoration bilevel model, but also prove the accuracy and efficiency of the proposed solution method, especially for the complex problems that are challenging for conventional methods.

The major contributions are summarized below:

\begin{itemize}[leftmargin=*]
    \item \textit{A Novel Method For Solving Bilevel Optimization Models}\\
    By representing the follower's behavior with a regressed function, the leader's omniscience assumption is not required. This approach owns a wider application range in practice and the higher computational efficiency than conventional solution methods, and can be used as a generic framework for various bilevel problems in the power system.

    \item \textit{A New Critical Load Restoration Bilevel Model}\\
    To the best of the authors' knowledge, this is the first time that the load-side operation and the varying MLV are considered together in power system restoration models. Compared to conventional methods, the developed model describes the load-side loss more accurately, and hence increase the optimality of restoration solutions.  
    
\end{itemize}
 The first contribution is to approximate the follower's optimal response. This approximation allows for the integration of the lower layer optimization into that of the upper layer, even in the absence of the leader's omniscience. The integration has a lower computational complexity compared to that of the conventional approach which converts the bilevel problem into a single layer one. As a result, the optimization problem of each layer can admit more complex, practical cost functions and constraints, which is impossible for the conventional approach, thus leading to the second contribution. 


\section{Bilevel Optimization Problem and Gaussian Process Regression}

\subsection{Bilevel Optimization Problems in the power system}
\begin{figure}[ht]
\centering 
\includegraphics[width=0.43\textwidth]{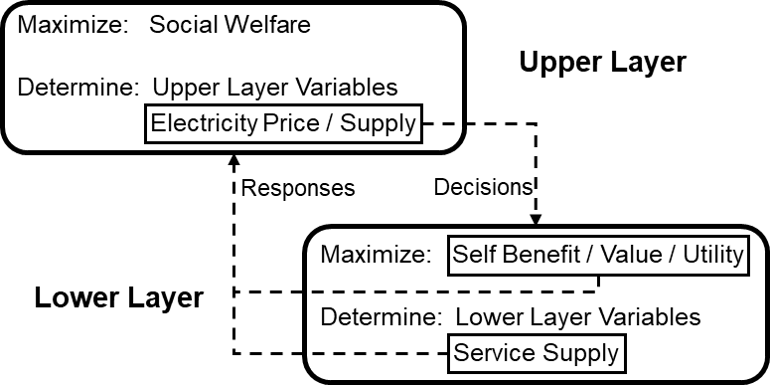}
\caption{A General Sketch of Bilevel Optimization Problems in the power system} 
\label{bilevel} 
\end{figure}

Bilevel optimization problems are designed for game leaders to optimize their decisions. The core concept of it is the decisions from a leader will affect the responses of the followers and then further affect the leader. In the power system, bilevel optimization problems can express the interaction between the power system and different entities. This interaction is collaborative in many cases, which the power system tends to maximize social welfare through purchasing power from resource owners and supplying it to customers as shown in Fig. \ref{bilevel}. For resource owners, the purchasing price will affect their trading strategy, and further affect the provided power amount. For customers, the amount of supplied power will affect their operation and behaviours, and further affect the worth and utility they made. Taking this interaction into account, the bilevel problem can be expressed as the following form.
\textcolor{black}{
\begin{align}
  & \min\limits_{d\in\mathbf{X_p},r\in\mathbf{X_f}}  F(d,r)   \label{leader's objective}\\
 \textrm{subject}&\;\textrm{to} \notag\\
& r\in \arg\min\limits_{r\in\mathbf{X_f}}\{f(d,r):g_j(d,r)\leq 0, j=1,...,J \} \label{follower's response}\\
& G_k(d,r) \leq 0, k=1,...,K \label{leader's constraints} 
\end{align}
}where $d$ represent the power system's decisions affecting the follower; $r$ represent the follower's responses taken into the power system' consideration; $\mathbf{X_p}$ is the power system operation variable set; $\mathbf{X_f}$ is the follower's operation variable set; $F$ and $f$ is the objective function of the power system and the followers, respectively; $G$ and $g$ represents the operation constraints of the power system and the followers, respectively. Obviously, for the the power system as the game leader running this model, they need to be omniscient to the followers' information to build the model.

\textcolor{black}{\subsection{Literature Review of Solution Methods for Bilevel Optimization Problems}}

Bilevel optimization is generally a NP-hard problem as the set of rational solutions of bilevel optimization problems is, in general, not convex and, in the presence of first-level constraints, may be empty (when upper-layer has no solution) \cite{NP-Hard1}. So classic solutions for bilevel problems usually assume the problems are mathematically well-behaved, such as linear or discrete linear models for both upper and lower layer \cite{linear_solution1,linear_solution2,linear_solution3}. 

\textcolor{black}{References \cite{single_level1} and \cite{single_level2} use the KKT conditions to replace the lower layer model, so that the bilevel model can be reduced to a single-level form. This approach is the most common solution method for the bilevel problem, but only suitable for the models with very specific forms. Another classic solution approach is the descent methods, which tend to find the descent direction decreasing the objective value of the upper layer model \cite{descent1,descent2}. Considering the feasible point of the upper layer must be the optimal solution of the lower layer, finding the descent direction is a challenging task, especially for the problems with strong interaction between the two layers. For the problems with less constraints, the penalty function methods may be very efficient. The penalty function methods remove the constraints in the original problem, and add penalty items in the objective function to describe the constraint violations \cite{penalty1,penalty2}. However, even the constraints are removed, the problem will still remain the bilevel form, which is difficult to be solved. Reference \cite{trust_region1} proposes a trust region method for bilevel problems, that a trust region of the objective function is approximated by algorithms with a model function. The size of the region is highly dependent on the approximation algorithms, which affects the universality of this type of methods.}

\textcolor{black}{Besides the classic approaches, some evolutionary methods are proposed in the recent decades. The evolutionary methods usually do not have strong assumptions on the problem formulation, but mostly likely incur a heavy computational burden, and thus most applicable to small-scale problems \cite{bilevel_review}. Popular evolutionary methods include the nested methods \cite{nest1,nest2}, evolutionary single-level reduction methods \cite{e_single_level1,e_single_level2}, and metamodeling-based methods \cite{metamodeling2}.
}

\textcolor{black}{Although a variety of approaches have been developed to solve bilevel optimization problems, most of them are based on the leader's omniscience assumption, which is not tenable in many scenarios of power system operation. Also, the computational efficiency is another concern which limits the application of many solution methods, especially when dealing with the large-scale problems. Therefore, an efficient solution method not relying on the leader's omniscience assumption is highly desired for the bilevel problems in power system operation.
}

\subsection{A Novel Solution Method based on Gaussian Process Regression}
\subsubsection{\textcolor{black}{Description of the Proposed Solution Method}}
In bilevel optimization problems, the followers' response $r$ is related to the power system's decision $d$. The original form of this relationship is the lower layer optimization problem shown in (\ref{follower's response}). If this optimization problem can be represented by a function with simpler form, the bilevel problem can be solved efficiently by the power system without knowing the follower's information. For some bilevel problems, it is difficult to obtain such function because the interaction between the leader and the followers may be very complex. But for the power system, this interaction is very straightforward: the decisions from the power system will be eventually reflected by the power generation or demand change of the followers, which makes it possible to represent the function with a single equation. To obtain the function, the power system should provide related information to the followers such as the varying range of $d$, and then the follower can generate the samples of $d$ and calculate the related response $r$. By doing the regression on the sample set, the representing function can be obtained and (\ref{follower's response}) is reformulated as
\begin{equation}
 r=h(d)\approx\hat{h}(d)
\end{equation}
where $h(\cdot)$ represents the relationship between the leader's decision $d$ and the follower's response $r$, and $\hat{h}(\cdot)$ is the estimated function of $h(\cdot)$ through regression. \textcolor{black}{It is worth noting that using an approximate function to represent the lower layer model does not change the interaction between the upper and lower layer. In other words, with the proposed solution method, the form of the bilevel problem is simplified, but the nature of the problem remains the same. Therefore, similar to the other solution approaches, the proposed method cannot guarantee the global optimal solutions for general bilevel problems.}

\subsubsection{\textcolor{black}{Key Assumptions of the Proposed Method}}

\textcolor{black}{The proposed solution method uses a regressed function to represent the lower layer model in a bilevel problem. Since the proposed method does not require certain forms of the lower layer model, the proposed method is a general solution approach for the bilevel problems in the power system. However, the following two assumptions should be satisfied to apply the proposed method.The first assumption is that some entities are not willing to share all information or control access with the power systems, even though they want to collaborate with the power system. In other words, there are information barriers between the power system and other entities. This assumption stands in many application scenarios, especially when the entity is not affiliated with the power system. An example is the IDC discussed in this paper. Some information in the IDC operation is confidential, such as the value of computing workloads \cite{abowd2004new}. Even though the IDC wants to collaborate with the power system to minimize its loss during the restoration process, these confidential information or control access cannot be revealed to avoid the attack from hackers. This assumption justifies the advantages of the proposed GPR based solution method. With the proposed solution method, the IDC only needs to provide a function describing the loss changing with the load shedding amount.}

\textcolor{black}{The second assumption in the paper is that the entities are willing to honestly describe how their optimal responses change with the power system's decisions. This assumption does not stand for all cases, especially when the entity is in a competitive position to the power system. However, it is reasonable to make that assumption in many collaborative scenarios, and an example is the critical load restoration discussed in this paper. Considering the power system is an operator of infrastructure with the objective to maximum social welfare, most load-side customers are willing to collaborate with the power system by reporting their actual loss during the restoration process. In fact, the questionnaires collected from customers is the main method of gathering information for the load-side loss evaluation \cite{billinton2005approximate,ahsan2004evaluation,wacker1989customer}. Moreover, even if an opportunistic presume wants to earn additional profit by providing false information, his behavior is also constrained by the competition among the followers. For example, a resource owner can bid any price he wants in an electricity market, but his optimal strategy is to bid according to the actual cost if a perfectly competitive market is assumed \cite{wang2016game}. 
}

\subsubsection{\textcolor{black}{Gaussian Process Regression}}
Although there are numbers of methods can be used for the regression work, there are many constraints on the regression method selection for this bilevel problem. Firstly, since Function $h(\cdot)$ essentially describes the relationship between two variables in an optimization problem, it is difficult to fit the function into certain forms. So the parametric and semi-parametric regression methods with predetermined function expressions cannot be used. Secondly, because the estimated function will be used as a part of the upper layer optimization model, an analytical expression of the function is required instead of a black box, which means most machine learning based approaches are out of consideration. Thirdly, since the lower layer problem may not be a convex optimization problem and the calculated responses in the generated samples may not be the optimal ones, which requires the regression method should consider sampling errors. 
Considering the above requirements, Gaussian Process Regression is selected for the regression work with the following reasons:
\begin{itemize}[leftmargin=*]
    \item  As a non-parametric regression method, GPR learns the form and the hyper-parameters of the regressed function from samples.
    \item GPR uses kernel functions to represent the covariance function. By selecting proper kernel functions, GPR can achieve good accuracy for various curves with limited number of samples.
    \item As a Bayesian regression method, GPR gives the confidence level of the estimation, which not only considers the estimation error, but also takes the sampling errors into account.
    \item Some non-parametric regression methods lead to large estimation biases near boundaries of variables, while GPR provides reliable estimation within the entire range of variables.
\end{itemize}

The procedures using GPR to estimate $r=h(d)$ is introduced here. Interested readers can refer to \cite{gpr} for more details. Defining $S=\left\{(d^{(i)},\hat{r}^{(i)}) \right\}_{i=1}^{m}$ is the sample set, where $\hat{r}^{(i)}$ is the observed value of $r$ with given $d$ at the $i^{th}$ sample. $\hat{r}^{(i)}$ can be expressed as:
\begin{align}
    & \hat{r}^{(i)}=h(d^{(i)})+\epsilon^{(i)}\\
    & \epsilon^{(i)}\sim N(0,\sigma^{(i)})
\end{align}
where $\epsilon^{(i)}$ is the error in the sample generation process and $\sigma^{(i)}$ is the standard deviation. By assuming the prior distribution of $h(\cdot)$ as a zero-mean Gaussian Process, the estimation of $h(d)$, or its mean value, with the given sample set $\hat{R}=[\hat{r}^{(1)},...,\hat{r}^{(m)}]^T$ at points $D=[d^{(1)},...d^{(m)}]$ is
\begin{equation} 
    \hat{h}(d)=k(d,D)^T(K(D,D)+\sigma^2I)^{-1}R \label{GPR mean function}
\end{equation}
with the standard deviation $\Sigma$ of $\hat{h}(d)$ at point $d$ is
\begin{equation}
    \Sigma(d)=k(d,d)+\sigma^2I-k(d,D)(k(D,D)+\sigma^2I)^{-1}k(D,d)
\end{equation}
where $k(\cdot,\cdot)$ is the kernel function. As shown in \eqref{GPR mean function}, the regressed function $\hat{h}(d)$ consists of the independent variable $d$, the kernel function, and constants $D$ and $R$ which are defined by the sample points. In GPR, kernel functions define the distance between two variable points, and the selection of kernel functions may affect the regression accuracy. Here, as most GPR applications did, the squared exponential (SE) kernel is used because of its universality as shown below,
\begin{equation}
    k_{SE}(x,x')=\sigma_{f}^2exp(-\frac{(x-x')^2}{2l^2})     
\end{equation}
where the scale factor $\sigma_{f}^2$ and lengthscale $l$ are the hyperparameters of SE kernel.

\section{Bilevel Critical Load Restoration Model Considering Lode-Side Operation and Varying Load Value}

\subsection{\textcolor{black}{Power System Restoration Problem Description}}
Post-event critical load restoration is critical to power system resilient operation. \textcolor{black}{The objective of the CL restoration is usually minimizing the loss of CLs due to interrupted power supply \cite{gao2016resilience}. The loss of CLs can be expressed as many forms, such as the economy loss and the utility loss, which usually can be expressed as a fixed cost coefficient $c_i$ times the load shedding amount as shown in Eq. (\ref{general_objective})  \cite{xu2016microgrids,xu2017dgs,lei2016mobile,arif2018optimizing,lei2019resilient}.} 

\textcolor{black}{
\begin{equation}
    \min \sum_{i} c_i*Load^{Shedding}_i \label{general_objective}
\end{equation}}

\noindent \textcolor{black}{Another mainstream definition of the restoration objective is to minimize the weighted load shedding amount as shown in Eq. (\ref{priority_weight}), where $p_i$ is the priority weight of the load $i$, which is assumed as a fixed value in most cases \cite{arif2018optimizing,lei2019resilient}.}

\textcolor{black}{
\begin{equation}
    \min \sum_{i} p_i*Load^{Shedding}_i \label{priority_weight}
\end{equation}}

\noindent \textcolor{black}{The above two restoration objectives share a same assumption that the utility of the load is directly proportional to the forced load shedding amount.}

However, this assumption may be not tenable in some cases. First, for some CLs, their workloads can be migrated temporally and spatially, and a typical example is the internet data center (IDC). During power supply interruption periods, the IDC can postpone computing workloads or transfer them to another IDCs, and even terminate the workloads with low value to minimize the loss. For these flexible loads like the IDC, their actual loss is not directly proportional to the interrupted power supply, but can be only obtained through the optimization of load-side operation \cite{liu_yang}. Second, customers' need for power is likely to follow the law of diminishing marginal returns, which means even for the non-flexible loads, the marginal load value decreases with power supply instead of a fixed value. This concept is widely accepted in demand response and load curtailment related research \cite{demand_response, load_curtailment}, but rarely used in restoration related problems. One reason is that although considering the two factors will increase modeling accuracy, it will introduce nonlinearity into the optimization problem and induce more computational burdens. 

\subsubsection{Upper Layer - Power System Restoration Model}  
For the power system restoration problem modeled in the upper layer, the objective function is:
\begin{equation}
\min\sum_{i}\sum_{t} C_i(P_{i,t}^d\,\rho_{i,t}^s) \label{upper_objective}
\end{equation}
where
\begin{align}
C_i(P_{i,t}^d\,\rho_{i,t}^s) &=a_i\left(P_{i,t}^d\,\rho_{i,t}^s\right)^3
+ b_i\left(P_{i,t}^d\,\rho_{i,t}^s\right)^2 \nonumber \\ &+ c_i \left(P_{i,t}^d\,\rho_{i,t}^s\right) \,\, \forall i\in\mathcal{B}\backslash \mathcal{B}^{idc},t \label{eq2}
\end{align}
\begin{align}
\sum_{i\in\mathcal{B}^{idc}}\sum_t C_i(P_{i,t}^d\,\rho_{i,t}^s)=L^{idc}=h\,(P_{i\in \mathcal{B}^{idc},t}^d\,,\rho_{i\in \mathcal{B}^{idc},t}^s)  \label{eq3}
\end{align}
The restoration objective is to minimize the load-side loss as shown in (\ref{upper_objective}), which is expressed as functions of load shedding amount for non-flexible and flexible loads as shown in (\ref{eq2}) and (\ref{eq3}). 

\begin{figure}[ht]
\centering 
\includegraphics[width=0.45\textwidth]{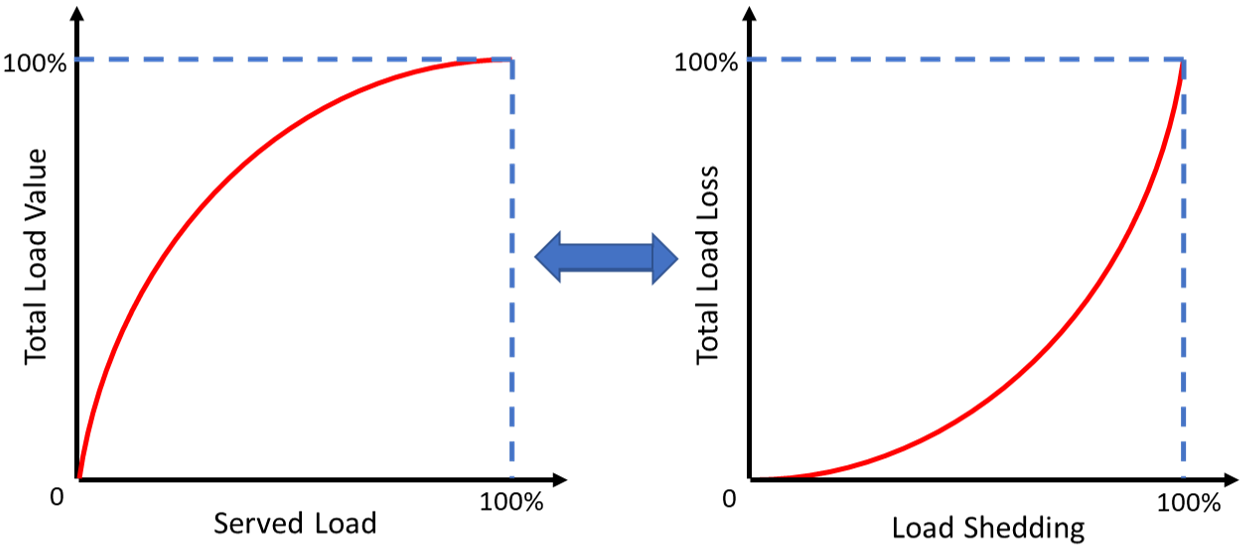} 
\caption{\textcolor{black}{Total Load Value and Loss v.s. Power Supply Amount}} 
\label{power_supply} 
\end{figure}

\textcolor{black}{For the non-flexible loads, their utility is directly proportional the load shedding. So the non-flexible load loss can be simply calculated based on the load shedding amount and directly contribute to the objective function in Eq. \eqref{upper_objective}. Considering the need of customers follows the law of diminishing marginal returns \cite{load_curtailment}, the relationship between the load shedding amount and the non-flexible load loss (or the total load value, oppositely) usually fits the shape shown in Fig. \ref{power_supply}.} This function can be represented by many forms based on historical data and questionnaire. Here, it is expressed as a cubic equation as shown in (\ref{eq2}), but other forms fitting the curve are also acceptable for this model. 

\textcolor{black}{For the flexible loads like the IDC, the load-side loss is not only determined by the load shedding amount, but also affected by the load-side operation. So the load-side operation model for the flexible loads will be built as the lower layer of the bilevel problem. From the power system point of view, the flexible load loss $L^{idc}$ is a function of the load shedding amount as shown in (\ref{eq3}), but the actual value of $L^{idc}$ can only be obtained through the load-side operation model in the lower layer.}

\begin{align}
&    \sum _{j}P_{i,j,t} = P_{i,t}^g-P_{i,t}^d(1-\rho_{i,t}^{s})    \quad  \forall (i,j)\in \mathcal{L},t \label{eq4}\\
&    \sum _{j}Q_{i,j,t} = Q_{i,t}^g-Q_{i,t}^d(1-\rho_{i,t}^{s})    \quad  \forall (i,j)\in \mathcal{L},t \label{eq5}\\
&    0\leq \rho_{i,t}^s \leq 1                                    \quad  \forall i,t \label{eq6}\\
&    \underline{P^g_{i}}\leq P^g_{i,t} \leq \overline{P^g_{i}}     \quad \forall i,t \label{eq7}\\
&    \underline{Q^g_{i}}\leq Q^g_{i,t} \leq \overline{Q^g_{i}}    \quad  \forall i,t \label{eq8}\\
&    \underline{V_{i}}\leq V_{i,t} \leq \overline{V_{i}}          \quad  \forall i,t \label{eq9}\\
& \begin{aligned} 
V_{j,t} \leq V_{i,t}-(r_{i,j}\,P_{i,j,t}+x_{i,j}\,Q_{i,j,t})+(1-u_{i,j,t})\,M \\
\forall (i,j)\in \mathcal{L},t  \label{eq10}
\end{aligned} \\
& \begin{aligned}
V_{j,t} \geq V_{i,t}-(r_{i,j}\,P_{i,j,t}+x_{i,j}\,Q_{i,j,t})-(1-u_{i,j,t})\,M \label{eq11}\\
\forall (i,j)\in \mathcal{L},t
\end{aligned} \\
&   P_{i,j,t}^2 + Q_{i,j,t}^2 \leq (S_{i,j}^{max})^2 \, u_{i,j,t}  \quad \forall (i,j)\in \mathcal{L},t \label{eq12}
\end{align}
\begin{align}
&   u_{i,j,t} \geq \frac{\sum_{t' \in [1,t-1]} u_{i,j,t'}^{r} - T_{i,j}^{r} +m}{M} \quad \forall (i,j)\in \mathcal{L^d},t \label{eq13}
\end{align}
\begin{align}
&   u_{i,j,t} \leq 1+\frac{\sum_{t' \in [1,t-1]} u_{i,j,t'}^{r} - T_{i,j}^{r}}{M} \quad \forall (i,j)\in \mathcal{L^d},t \label{eq14}\\
&   \sum_{(i,j)\in \mathcal{L^d}}u_{i,j,t}^{r} \leq N^R        \quad \forall t \label{eq15} \\
&   u_{i,j,t} \leq 1-u_{i,j}^{sw}                       \quad \forall (i,j)\in \mathcal{L}\backslash\mathcal{L^d} \label{eq16}
\end{align}
The operation of the power system during restoration periods are modeled in (\ref{eq4}) - (\ref{eq16}). \textcolor{black}{Considering the power network connected to the IDC is usually in distribution level, the restoration model is designed for distribution systems \cite{geng2014data}. The active and reactive power balance in the network is modeled in \eqref{eq4} and \eqref{eq5}. The load shedding percentage limits is formulated in \eqref{eq6}. Equation \eqref{eq7} and \eqref{eq8} describe the active and reactive power generation limits. The voltage magnitude limits is modeled in \eqref{eq9}, and the power flow in the network is formulated with DistFlow equations as shown in \eqref{eq10} and \eqref{eq11} \cite{baran1989optimal}. The transmission limits of power lines are modeled in \eqref{eq12}. Equation (\ref{eq13}) and (\ref{eq14}) reflect that the damaged line will be repaired only after the total repairing time is equal or larger than the required time, and (\ref{eq15}) represents that the number of simultaneously repairing lines should not exceeds the number of repairing crews. Equation \eqref{eq16} shows that the line equipped with remotely controlled switches can be used for network reconfiguration during restoration.}

\subsubsection{Lower Layer - Load-Side (Internet Data Center) Operation Model}
The objective function of the load-side operation modeled in the lower layer is:
\begin{equation}
   \min L^{idc}=\sum_{wl}W_{wl}\,v^t_{wl}  \label{eq16.5}
\end{equation}

The operation objective of flexible loads during restoration periods is to minimize the loss caused by terminating the workloads which are originally scheduled to be completed as shown in (\ref{eq16.5}). The termination of workloads $v^t_{wl}$ in the IDC is affected by the IDC load shedding amount $\rho_{i\in \mathcal{B}^{idc},t}^s$, which is determined by the power system restoration model in the upper layer. 

The list of constraints at the lower level describing the operation of the IDC are modeled in \eqref{eq17} - \eqref{eq24} below. Equation (\ref{eq17}) describes a workload will be terminated if it is not completed in either IDC. Equation (\ref{eq18}) and (\ref{eq19}) indicate that a workload is completed when the allocated server resource amount between its release time and its deadline is larger than or equals to the required amount. Equation (\ref{eq20}) shows all workloads can be only completed once. For each time period, the total allocated server resource in a IDC should not be larger than the server capacity of the IDC as shown in (\ref{eq21}). Equation (\ref{eq22}) - (\ref{eq24}) describes the workload migration among IDCs. Originally, a workload can be only computed in the IDC where it locates. But after a workload is uploaded to cloud from its original location, the workload can be computed by other IDCs. The number of uploaded workloads from a IDC should not exceeds its uploading limit. Compared to the previous work \cite{liu_yang}, %
the migration modeling is simplified based on the actual operation in the IDC and can significantly reduce the number of binary variables in the model.
\begin{align}
&   v^t_{wl}=1-\sum_{idc}v^c_{wl,idc}  \quad \forall wl \label{eq17}\\
&   v^c_{wl,idc} \geq \frac{(\sum_{t\in [RT_{wl},DL_{wl}]}S_{wl,idc,t})-S_{wl}^{req}+m}{M} \quad \forall wl,idc \label{eq18}\\
&   v^c_{wl,idc} \leq 1-\frac{S_{wl}^{req}-\sum_{t\in [RT_{wl},DL_{wl}]}S_{wl,idc,t}}{M}   \quad \forall wl,idc \label{eq19}\\
&   \sum_{idc}v^c_{wl,idc} \leq 1       \quad \forall idc \label{eq20} \\
&  0 \leq \sum_{wl}S_{wl,idc,t} \leq S_{idc}^{cap} \quad \forall idc,t \label{eq21} \\
&  S_{wl,idc,t} \leq (v^l_{wl,idc} + \sum_{idc} \sum_{t'\in[1,t]} v^u_{wl,idc,t'})\,M \quad \forall wl,t \label{eq22} \\
&  v^u_{wl,idc,t} \leq v^l_{wl,idc}    \quad \forall wl,idc,t  \label{eq23} \\
&  \sum_{wl}v^u_{wl,idc,t} \leq UP^{max}_{idc} \quad \forall t \label{eq24} \\
& \begin{aligned}
 \sum_{idc\in\mathcal{IDC_i}}\sum_{wl}S_{wl,idc,t}\,(1+1/LF_{idc,t}) = P_{i,t}^d\,(1-\rho_{i,t}^s) \\
\forall i\in\mathcal{B}^{idc},t \label{eq25}
\end{aligned}
\end{align}

The influence of the upper layer's decisions to the lower layer is reflected by (\ref{eq25}). The left side of (\ref{eq25}) is the power demand of the IDC, which is computed based on the allocated server resource amount and a load factor $LF_{idc,t}$ describing the ratio between the energy consumed by servers and cooling devices \cite{li2011towards}. The right side of (\ref{eq25}) is the power demand can be satisfied determined by the upper layer, which will affect the available server resource in the IDC, and further affect the IDC load-side loss.


\subsubsection{\textcolor{black}{Bilevel Restoration Model vs Conventional Single-level Restoration Model}}

\textcolor{black}{In the above model, the objective of the lower layer directly contributes to the upper layer model, which means the proposed bilevel restoration model can be merged as a single-level integrated model. In fact, most of existing approaches introduced in Section III.B describe the power system restoration problem as a single-level model. The justifications of the bilevel restoration model is provided below.}

\textcolor{black}{Firstly, for the power system restoration, the power system tends to minimize the total load-side loss during the restoration process. The restoration scheme designed by the power system will affect the behaviors of the customers, which will determine the load-side loss. By forming the restoration problem as a bilevel form, the interaction between the power system and the load-side customers can be clearly formulated.} 

\textcolor{black}{Secondly, the actual power restoration in reality may be a multi-objective problem. Besides minimizing the load-side loss, other concerns such as the number of switch operations and the downgrade of ESS may also be optimized in the restoration scheme designing \cite{kumar2007multiobjective}. Similarly, minimizing the loss caused by the power supply interruption may not be the only consideration for the load-side customers. For example, the IDC may dispatch all computing workloads to one server and shut down other servers to increase the energy usage efficiency, so that more workloads can be completed during the power shortage periods. However, concentrating workloads on single server may lead to heat imbalances and create hot spots, and further affect the life of servers and AC systems \cite{xu2010multi}. Therefore, the objectives of the power system and the load-side customers are not same in many cases, and the bilevel problem cannot be always merged to the single level model.}

\textcolor{black}{Last but not least, modeling the restoration problem as a single level form is based on the assumption that the power system knows all information of the load-side customers. As discussed in Section II.B, this leader omniscience assumption does not stand in all cases. When the load-side customers do not want to share all operation details, the single-level restoration model cannot be built.}

\subsection{Solving the Bilevel Critical Load Restoration Model with GPR-based Solution Method}

\begin{figure}[ht]
\centering 
\includegraphics[width=0.45\textwidth]{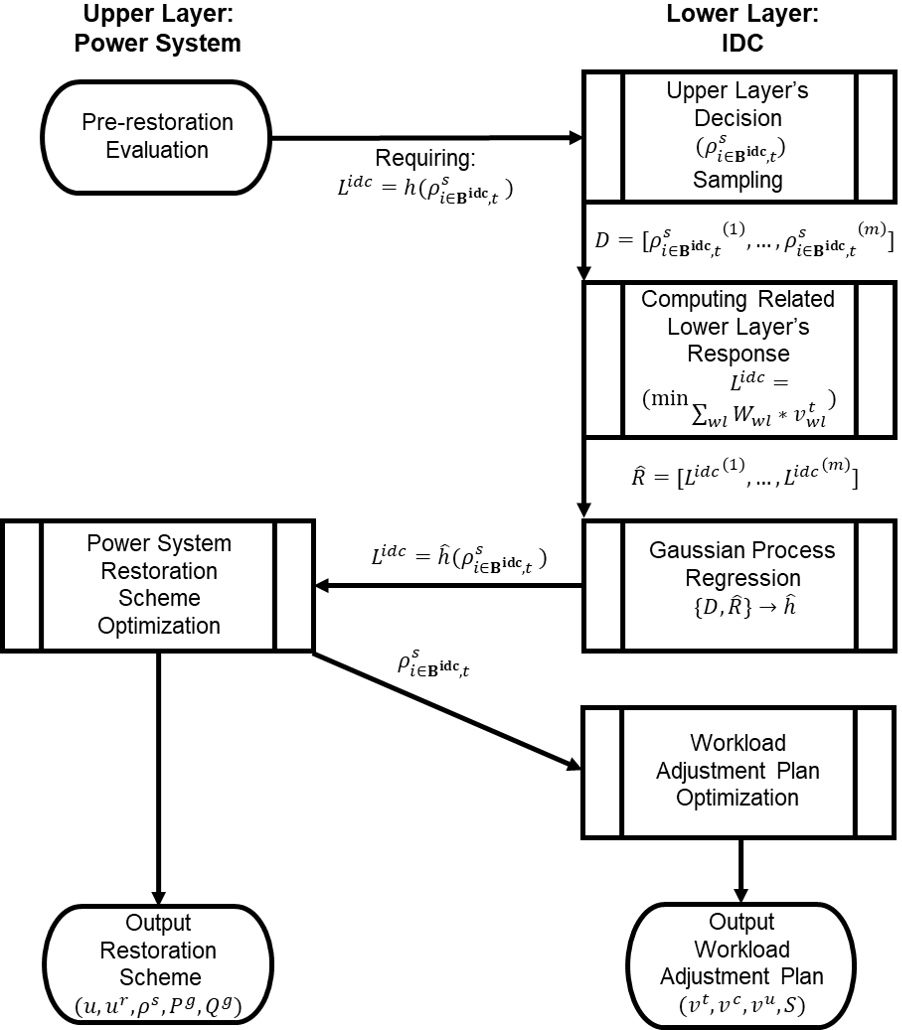}
\caption{Procedures of Solving Bilevel Critical Load Restoration Model with GPR based Solution Method} 
\label{procedures} 
\end{figure}

To solve this restoration model, the classical way is to reduce it to a single level problem by replacing (\ref{eq3}) with (\ref{eq16.5}) and merging the upper and lower layer constraints. However, it becomes difficult when the lower layer is unaware by the upper layer. The proposed GPR based solution method can overcome this issue and solve the model, and the detailed procedures are shown in Fig. \ref{procedures}. 

In the bilevel restoration model, the upper layer's decisions on the IDC load shedding ($\rho^s_{i\in\mathcal{B}^{idc},t}$) will affect the IDC' loss ($L^{idc}$) in the lower layer. So the function $h(\cdot)$ in (\ref{eq3}) describing the IDC's loss and IDC load shedding will be required by the power system to design restoration schemes. To estimate $h(\cdot)$, the IDC will generate a sample set $D$ of $\rho^s_{i\in\mathcal{B}^{idc},t}$, and then calculate ($L^{idc}$) to build the response set $R$. During sampling, a GP-UCB based sampling method \cite{zhaichao,pareek2020probabilistic} will help to minimize the sample number while maintaining the required confidence level. After the sampling data is obtained, the estimated function $\hat{h}(\cdot)$ will be obtained through GPR as expressed in (\ref{GPR mean function}). By taking $\hat{h}(\cdot)$ to (\ref{eq3}), the restoration model in the upper layer can be solved by the power system without knowing the details of lower layer. Then the load shedding decisions $\rho^s_{i\in\mathcal{B}^{idc},t}$ will be informed to the IDC to optimize the load-side operation.

\section{Case Studies}
Here the proposed GPR-based solution method will be used to deal with two power system restoration cases, which represent two types of bilevel problems with different complexity and size. The performance of the proposed restoration model and the GPR-based solution method will be analyzed and compared to conventional methods. The optimization models in the case studies are solved through the solver BARON running in a PC with 3.2GHz CPU and 16GB RAM. 

\subsection{Single-Period Critical Load Restoration with Fixed Marginal Load Value in Modified IEEE 33-bus System}
\begin{figure}[ht]
\centering 
\includegraphics[width=0.48\textwidth]{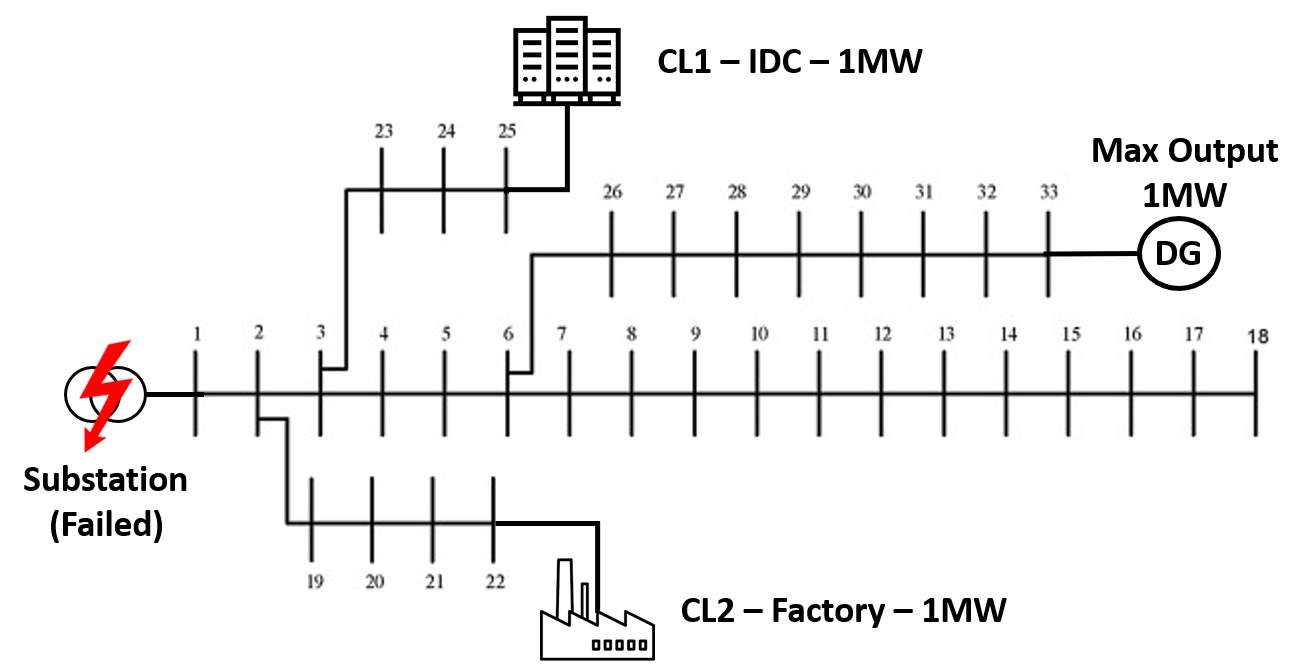} 
\caption{\textcolor{black}{Modified IEEE 33 Bus Distribution System with One Internet Data Center and One Factory as Critical Loads}} 
\label{case33} 
\end{figure}

In the first case, two CLs and one distributed generator (DG) are assumed connecting to IEEE 33-bus system as shown in Fig. \ref{case33}. The information of the CLs is shown in Tab. \ref{tab1}. CL1 is an internet data center with flexible workloads, and CL2 is a factory - a typical non-flexible CL. The main substation is assumed to be failed, and the DG will be used to restore the two CLs. Since the maximum DG's output is less than the CLs' power demand, the key of the restoration is to optimally allocate the power supply to the two CLs.  

The purpose of this case study is to i) test the performance of the GPR-based method on a simple bilevel model, and ii) analyze how the load-side operation affects the restoration. For the flexible load IDC, its loss is determined by the load-side operation modeled in the lower layer; for the non-flexible load - the Factory and other normal loads, its loss is linearly related to power supply amount (varying MLV is not considered in this case). Since line reparations and network reconfigurations are not required in this case, the power system restoration modeled in the upper layer is a linear programming (LP) problem, and the IDC-side operation modeled in the lower layer is a mixed integer linear programming (MILP) problem.

\begin{table}[ht]
\centering
\caption{Information of Critical Loads}
\begin{tabular}{ccccc}
\toprule
 & Load Type & \begin{tabular}[c]{@{}c@{}}Marginal \\ Load Value\end{tabular} & Rated Power & \begin{tabular}[c]{@{}c@{}}Total \\ Load Value\end{tabular}\\
\midrule
CL1& IDC      & Workload Based & 1MW & \$5,000k\\
CL2& Factory & Fixed          & 1MW & \$6,000k\\
\bottomrule
\end{tabular}
\label{tab1}
\end{table}

\noindent \textcolor{black}{\textit{\textbf{Applying the Proposed GPR based Solution Method on the Case Study}}}\\
\hspace*{0.5cm} \textcolor{black}{The case study is modeled as the bilevel restoration model with the upper layer as the power system restoration model Eq. (\ref{upper_objective}) - (\ref{eq16}) and the lower layer as the IDC operation model Eq. (\ref{eq16.5}) - (\ref{eq25}). The proposed GPR-based solution method is applied to solve the bilevel model following the procedures shown in Fig. \ref{procedures}. After the substation fails, the power system requires the IDC to provide the function $L^{idc}=h(\rho^s_{i\in\mathcal{B}^{idc},t})$ to describe the IDC's loss. To obtain $h(\rho^s_{i\in\mathcal{B}^{idc},t})$, the IDC generates multiple samples of the possible value of $\rho^s_{i\in\mathcal{B}^{idc},t}$, and calculates the related $L^{idc}$ for each sampled $\rho^s_{i\in\mathcal{B}^{idc},t}$ through Eq. (\ref{eq16.5}) - (\ref{eq25}) to build the sample set $[D,\hat{R}]$. Then the sample set is inputed to Eq. (\ref{GPR mean function}) to regress the function $L^{idc}=\hat{h}(\rho^s_{i\in\mathcal{B}^{idc},t})$. The regressed function is provided to the power system as Eq. (\ref{eq3}), and then the optimal restoration scheme and the related load-side loss can be obtained by only solving the upper layer model with the regressed function as shown below:
\begin{align}
    &\textrm{Objective:} \quad\textrm{Eq.}\: \eqref{upper_objective} \label{eq26}\\ 
    &\textrm{Subject to} \quad\textrm{Eq.}\: \eqref{eq2},\eqref{eq4} - \eqref{eq16},\:L^{idc}=\hat{h}(\rho^s_{i\in\mathcal{B}^{idc},t}) \label{eq27}
\end{align}
For comparison, the conventional solution method is also used to solve the bilevel model. The conventional solution method merged the bilevel model into an integrated single-level model as shown below: 
\begin{align}
    &\textrm{Objective:} \quad\textrm{Eq.}\: \eqref{upper_objective} \label{eq28}\\
    &\textrm{Subject to} \quad\textrm{Eq.}\: \eqref{eq2},\eqref{eq4} - \eqref{eq16},\:\eqref{eq16.5} - \eqref{eq25} \qquad\quad \label{eq29}
\end{align}
}
\textcolor{black}{It can be seen that the major difference between the proposed solution method and the conventional solution method is that the lower layer model in Eq. \eqref{eq16.5} - \eqref{eq25} is represented by the function $\hat{h}(\cdot)$. The performance change brought by this difference will be analyzed in the following parts.
}\\

\noindent \textit{\textbf{Simulation Results and Solution Method Performance}}\\
\hspace*{0.5cm}The restoration results using the bilevel CL restoration model are shown in Tab. \ref{table.case33_restoration}, and compared to the conventional restoration model that treats the IDC as a non-flexible load without load-side operation. With the conventional model, all power supply is allocated to the factory because its total value after fully restoration is higher as shown in Tab. \ref{tab1}. On the other side, with the proposed model considering load-side operation, 40\% of the power supply is allocated to the IDC and the total loss is much lower than the conventional model.

\begin{table}[ht]
\centering
\caption{Restoration Results w/wo Considering IDC-Side Operation} 
\label{table.case33_restoration}
\begin{tabular}{cccccc}
\toprule
\begin{tabular}[c]{@{}c@{}}Restoration\\ Model\end{tabular} &
  \begin{tabular}[c]{@{}c@{}}IDC\\ Power\\ Supply\\ (MW)\end{tabular} &
  \begin{tabular}[c]{@{}c@{}}Factory\\ Power\\ Supply\\ (MW)\end{tabular} &
  \begin{tabular}[c]{@{}c@{}}IDC\\ Loss\\ (\$)\end{tabular} &
  \begin{tabular}[c]{@{}c@{}}Factory\\ Loss\\ (\$)\end{tabular} &
  \begin{tabular}[c]{@{}c@{}}Total\\ Loss\\ (\$)\end{tabular} \\
\midrule
\begin{tabular}[c]{@{}c@{}}Consider\\ IDC-Side\\ Operation\end{tabular} &
  0.40 &
  0.60 &
  1,830k &
  2,400k &
  \begin{tabular}[c]{@{}c@{}}4,230k\\ (Optimal)\end{tabular} \\
\midrule
\begin{tabular}[c]{@{}c@{}}Not Consider\\ IDC-Side\\ Operation\end{tabular} &
  0 &
  1 &
  5,000k &
  0 &
  5,000k\\
\bottomrule
\end{tabular}
\end{table}

The performance of the proposed GPR-based solution method is shown in Tab. \ref{table.case33_efficiency}. It can be seen that although representing the lower layer as a regression function introduces non-linearity into the problem, the problem can still be solved within one second. \textcolor{black}{Using the conventional solution method guaranteeing global optimal solutions as the benchmark, the error of the proposed method, which is defined as the deviation of the objective from the global optimal value, is smaller than 0.1\%, proving the GPR-based solution method is reliable.}\\ 

\begin{table}[ht]
\centering
\caption{\textcolor{black}{Performance of Proposed and Conventional Solution Method}}
\label{table.case33_efficiency}
\begin{tabular}{c|cccc}
\toprule
\multicolumn{2}{c}{\begin{tabular}[c]{@{}c@{}}Solution\\ Method\end{tabular}} &
  \begin{tabular}[c]{@{}c@{}}Problem\\ Type\end{tabular} &
  \begin{tabular}[c]{@{}c@{}}Variable\\ Number\end{tabular} &
  Performance \\
\midrule
\begin{tabular}[c]{@{}c@{}}Proposed\\ Method\end{tabular} &
  \begin{tabular}[c]{@{}c@{}}Upper\\ Lower\\ Combined\end{tabular} &
  \begin{tabular}[c]{@{}c@{}}LP\\ NLP\\ NLP\end{tabular} &
  \begin{tabular}[c]{@{}c@{}}260\\ 1\\ 261\end{tabular} &
  \begin{tabular}[c]{@{}c@{}}0.7s\\ Error \textless 0.1\%\end{tabular} \\
\midrule
\begin{tabular}[c]{@{}c@{}}Conventional\\ Method\end{tabular} &
  \begin{tabular}[c]{@{}c@{}}Upper\\ Lower\\ Combined\end{tabular} &
  \begin{tabular}[c]{@{}c@{}}LP\\ MILP\\ MILP\end{tabular} &
  \begin{tabular}[c]{@{}c@{}}260\\ 201\\ 461\end{tabular} &
  \begin{tabular}[c]{@{}c@{}}0.1s\\ Benchmark\end{tabular}\\
\bottomrule
\end{tabular}
\end{table}

\begin{figure}[ht]
\centering 
\includegraphics[width=0.48\textwidth]{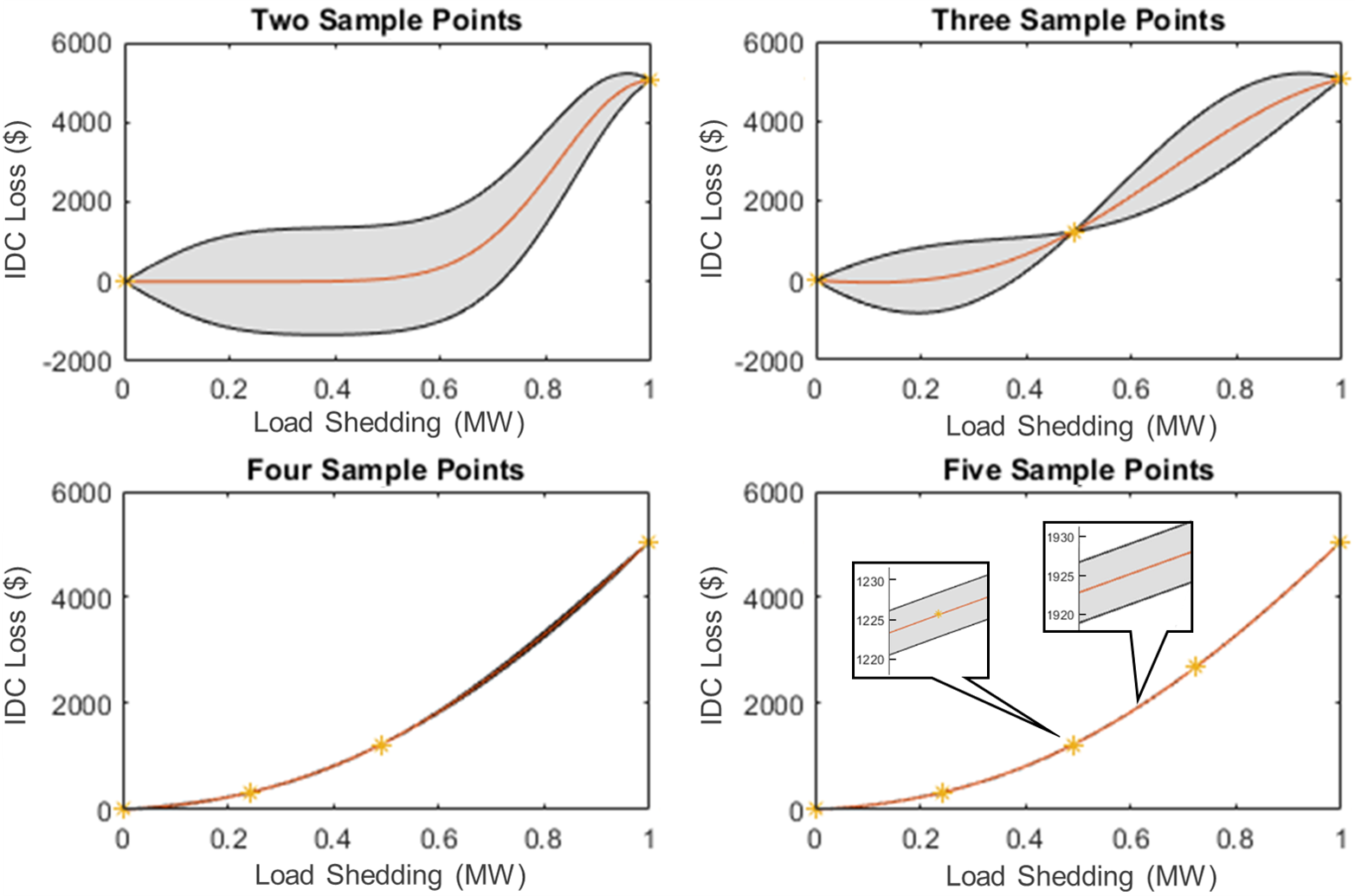} 
\caption{\textcolor{black}{Confidence level increasing with sample points}} 
\label{case33_sample} 
\end{figure}

\noindent \textit{\textbf{Discussion: The Influence of Regression Samples on GPR-based Method}}\\
\hspace*{0.5cm} 
In the GPR-based method, an important factor affecting its performance is the samples for regression. A good sample set can guarantee the accuracy of the regression and the reliability of the obtained optimal solutions. As aforementioned, one advantage of GPR is that it provides the confidence level of the regression result, which allows us to add samples at the inconfident points. \textcolor{black}{Fig. \ref{case33_sample} shows the processing of adding sample points according to the confidence level (the grey area). It can be seen that the regression variance significantly decreases with the increasing number of samples. After the fifth sample is added, although the variance between two sample points is still slightly higher than the variance at the sample points as shown in the fourth graph of Fig. \ref{case33_sample}, the unknown function is accurately regressed based on the five samples.}

\begin{figure}[ht]
\centering 
\includegraphics[width=0.42\textwidth]{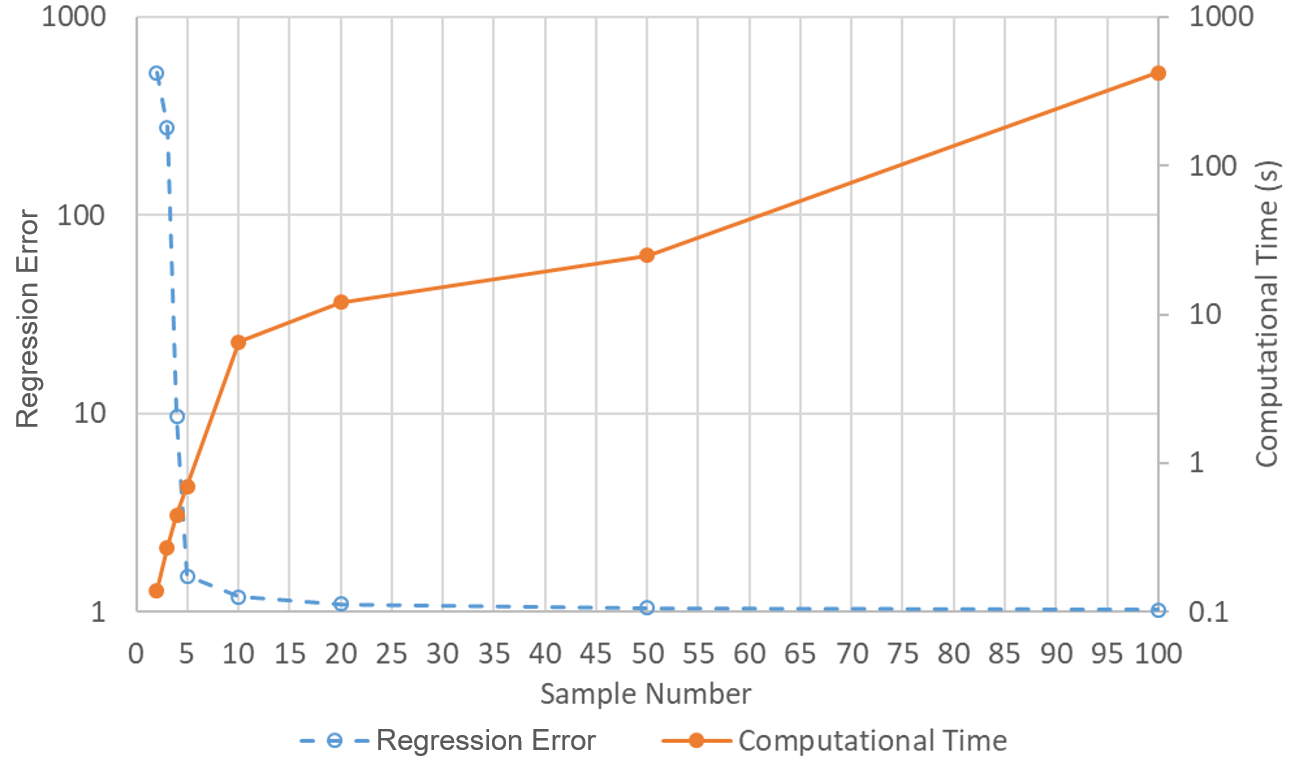} \caption{Elapsed Time and Regression Error Varying with the Number of Samples} 
\label{case33_computational_time} 
\end{figure}

Moreover, the regression samples also affect the computation of the proposed method. Eq. \eqref{GPR mean function} shows that the length of $\hat{h}(\cdot)$ will increase with the number of samples, which will further increase the elapsed time. \textcolor{black}{Fig. \ref{case33_computational_time} shows how the computational time and regression error varies with the number of samples (the error is defined as the absolute deviation between the original function and the regressed function).} Therefore, a conclusion can be drawn that do not add more samples if the regression accuracy is satisfying.

\noindent \textit{\textbf{\textcolor{black}{Discussion: Regression Accuracy and Efficiency of GPR}}}\\
\hspace*{0.5cm} 
\textcolor{black}{
As introduced in Section II.C, the performance of GPR is affected by the selection of kernel functions, which may further affect the flexibility of the proposed GPR based solution method. Therefore, another simulation is designed to test the regression accuracy and efficiency of GPR with different kernel functions. In this simulation, the GPR is used to regress the randomly generated unknown functions with the similar shape as the curve in Fig. \ref{power_supply} based on five samples. According to the simulation results shown in Tab. \ref{table.different_kernels}, the performance of GPR based mainstream kernel functions is stable. No matter which kernel function is selected, the regression can be completed within 1 second, and the regression error (defined as the deviation between the regressed value and the actual value in the percentage form) is slightly higher than 2\%. Compared to the two benchmark methods which are widely used for nonlinear regression, the performance of GPR is satisfying and reliable.
}

\begin{table}[ht]
\centering
\caption{\textcolor{black}{Regression Accuracy and Efficiency of GPR}}
\label{table.different_kernels}
\begin{tabular}{c|cccc}
\toprule
\multicolumn{2}{c}{\multirow{2}{*}{Method}} &
  \multirow{2}{*}{\begin{tabular}[c]{@{}c@{}}Consumed \\ Time\end{tabular}} &
  \multicolumn{2}{c}{Regression Error} \\

\multicolumn{2}{c}{} &
   &
  \begin{tabular}[c]{@{}c@{}}Within \\ Sample\end{tabular} &
  \begin{tabular}[c]{@{}c@{}}Out of \\ Sample\end{tabular} \\
\midrule
\multirow{5}{*}{GRE} & SE Kernel                 & \textless{}1s & 1.84\% & 2.17\% \\
                     & Exponential Kernel        & \textless{}1s & 1.86\% & 2.22\% \\
                     & Matern 3/2 Kernel         & \textless{}1s & 1.83\% & 2.17\% \\
                     & Matern 5/2 Kernel         & \textless{}1s & 1.83\% & 2.17\% \\
                     & Rational Quadratic Kernel & \textless{}1s & 1.84\% & 2.17\% \\
\midrule
\multirow{2}{*}{BM}  & Regression Tree           & 2s            & 1.88\% & 2.21\% \\
                     & Neural Network            & 10s           & 0\%    & 6.55\% \\
\bottomrule
\end{tabular}
\end{table}

\begin{figure}[ht]
\centering 
\includegraphics[width=0.45\textwidth]{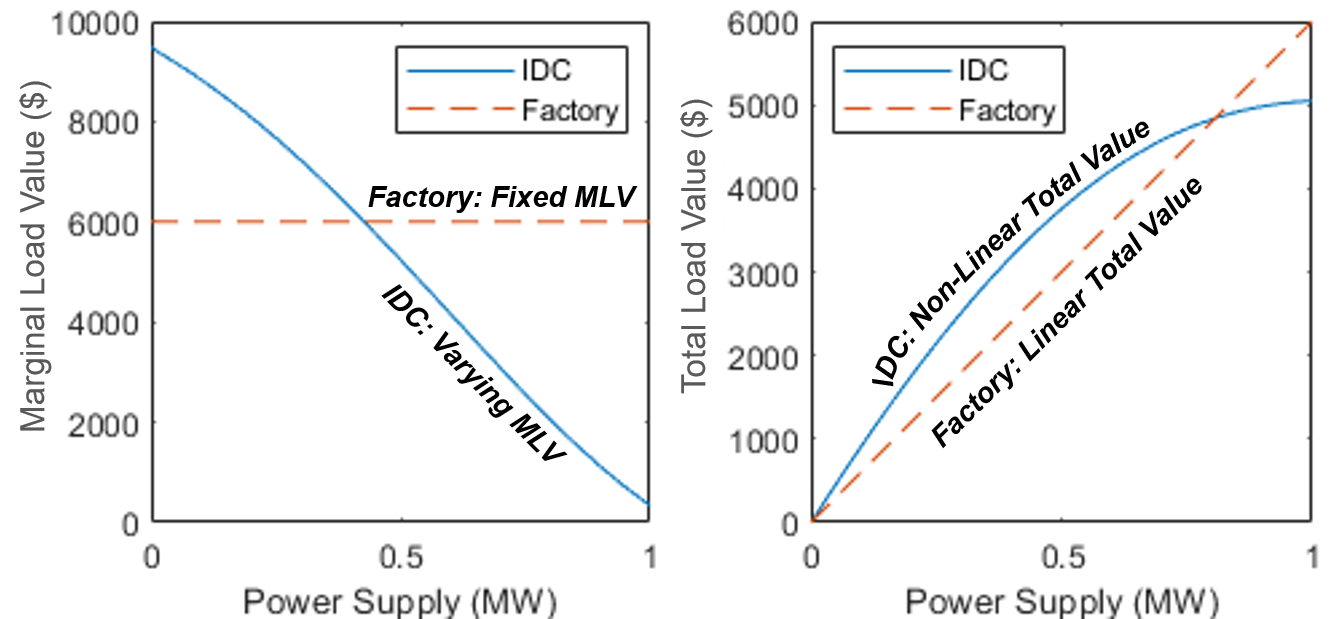} 
\caption{Marginal and Total Load Value of IDC and Factory Varying with Available Power Supply Amount} 
\label{case33_utility} 
\end{figure}

\noindent \textit{\textbf{Discussion: Considering Load-Side Operation in the Restoration Model}}\\
\hspace*{0.5cm}
The importance of considering load-side operation in restoration is discussed here. When facing power supply shortage, the IDC tend to complete the workloads with high values first, so the marginal value of the power supply is very high at first, and then decreases with the power supply increase as shown in the left part of Fig. \ref{case33_utility}. Compared to the IDC with flexible loads, the MLV of non-flexible loads such as the factory is a fixed value. Fig. \ref{case33_utility} shows that the MLV of IDC is larger than the factory when the power supply is smaller than 0.4 MW. Therefore, although the total load value of the factory with fully power supply is higher than the IDC, the optimal solution should be allocating 0.4 MW power to the IDC and the rest to the factory. 

\subsection{Multiple-Period Distribution System Restoration with Varying Marginal Load Value in Modified IEEE 123-bus System}
\begin{figure}[H]
\centering 
\includegraphics[width=0.48\textwidth]{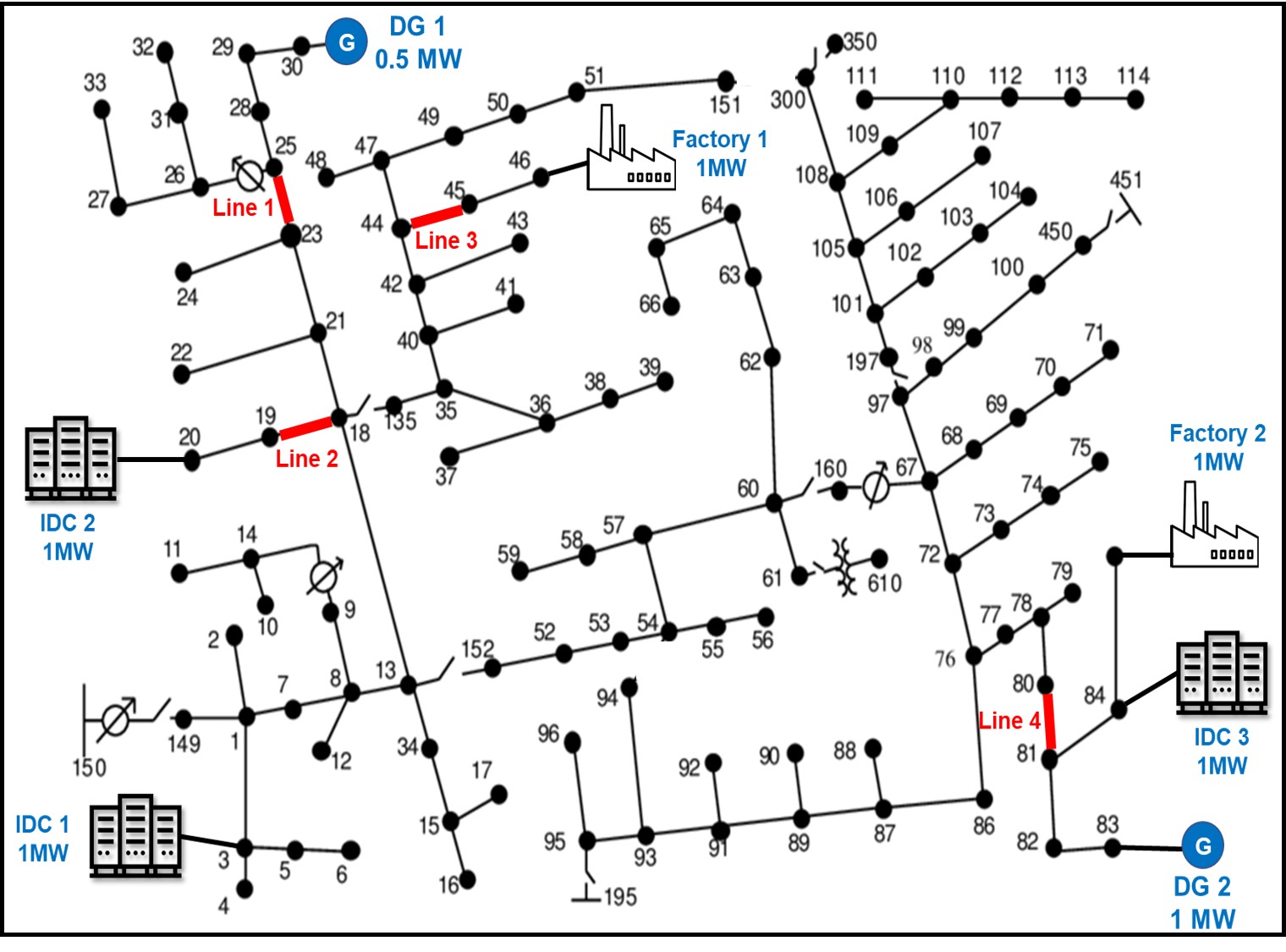} 
\caption{Modified IEEE 123 Bus Distribution System with Five Critical Loads and Two Distributed Generator} 
\label{case123} 
\end{figure}

In this case study, the performance of the GPR-based solution method on a complex bilevel problem will be tested, and the influence of the varying MLV of non-flexible loads will be analyzed. The case is built based on IEEE 123-Bus system with five CLs and two DGs, and four distribution lines are assumed to be failed as shown in Fig. \ref{case123}. It is assumed only one repairing crew is available and the repairing time of damaged lines is the same, so the restoration process is divided into four time periods. The system operator needs to decide the repairing sequence of the damaged lines and the generation dispatch scheme to restore the lost loads.

Different with the previous case, the varying MLV of non-flexible loads is considered, as well as the line reparation and network reconfiguration, which makes the restoration model become a mixed integer non-linear programming (MINLP) problem. The total load value and power demand of CL2-CL5 are assumed as the same as shown in Tab. \ref{table.case123_cl_information}, which makes it challenging to determine the restoration priority. For CL1-CL3 as IDCs, each IDC is expected to compute 80 workloads with different value, and the workloads are allowed to migrate among IDCs. The workloads in IDC1 and IDC2 can be completed at any time slots, but the workloads in IDC3 are constrained by release time and deadlines. 

\begin{table}[H]
\centering
\caption{Information of Critical Loads}
\label{table.case123_cl_information}
\begin{tabular}{ccccc}
\toprule
 & Load Type & \begin{tabular}[c]{@{}c@{}}Marginal \\ Load Value\end{tabular} & \begin{tabular}[c]{@{}c@{}} Power \\ Demand\end{tabular} & \begin{tabular}[c]{@{}c@{}} Total \\ Load Value\end{tabular}\\
\midrule
CL1& IDC1      & Workload Based & 1MW & \$800k\\
CL2& IDC2      & Workload Based & 1MW & \$2,400k\\
CL3& IDC3      & Workload Based & 1MW & \$2,400k\\
CL4& Factory 1 & Power Supply Related          & 1MW & \$2,400k\\
CL5& Factory 2 & Power Supply Related          & 1MW & \$2,400k\\
\bottomrule
\end{tabular}
\end{table}

\vspace*{0.5\baselineskip} 
\noindent \textit{\textbf{Simulation Results and Solution Method Performance}}\\
\hspace*{0.5cm}
Here except the proposed model, two other methods are used to find restoration solutions for comparison: the bilevel model with load-side operation but not considering the varying MLV of non-flexible loads and the conventional model not considering load-side operation or varying MLV. Table \ref{table.case123_result} shows the restoration results with the three methods.  

\begin{table}[ht]
\centering
\caption{Restoration Results with Different Methods}\label{table.case123_result}
\begin{tabular}{ccccc}
\toprule
 &
  \begin{tabular}[c]{@{}c@{}}Line\\ Reparation\end{tabular} &
  \begin{tabular}[c]{@{}c@{}}IDC\\ Loss\end{tabular} &
  \begin{tabular}[c]{@{}c@{}}Factory\\ Loss\end{tabular} &
  \begin{tabular}[c]{@{}c@{}}Total\\ Loss\end{tabular} \\
\midrule
\begin{tabular}[c]{@{}c@{}}Proposed\\ Model\end{tabular} &
  \begin{tabular}[c]{@{}c@{}}Line 3 $\rightarrow$ Line 2 $\rightarrow$\\ Line 4 $\rightarrow$ Line 1\end{tabular} &
  \$1,711k &
  \$958k &
  \begin{tabular}[c]{@{}c@{}}\$2,669k\\ (Optimal)\end{tabular} \\
\midrule
\begin{tabular}[c]{@{}c@{}}Only Consider\\ Load-side \\ Operation \end{tabular} &
  \begin{tabular}[c]{@{}c@{}}Line 3 $\rightarrow$ Line 2 $\rightarrow$\\ Line 4 $\rightarrow$ Line 1\end{tabular} &
  \$1,746k &
  \$1,206k &
  \$2,952k \\
\midrule
\begin{tabular}[c]{@{}c@{}}Conventional\\ Model\end{tabular} &
  \begin{tabular}[c]{@{}c@{}}Line 4 $\rightarrow$ Line 2 $\rightarrow$\\ Line 3 $\rightarrow$ Line 1\end{tabular} &
  \$1,250k &
  \$1,800k &
  \$3,080k \\
\bottomrule
\end{tabular}
\end{table}

Apparently, the proposed restoration model achieves the best restoration solution among the three methods. For the conventional method, since it cannot rank the priority of the four CLs with same power demand and total value, the line repairing sequence is determined based on the restored normal load, and therefore leads to the biggest loss. For the second method considering load-side operation and fixed MLV, the obtained line repairing sequence is the same as the proposed method, but result in higher total loss. This is because of the different generation dispatch plans and will be illustrated later.  

Similar to the first case, the restoration model is solved by the proposed GPR-based method and the conventional single-level reduction method. \textcolor{black}{For the GPR-based method, the function representing the lower layer is regressed based on 1,000 samples to show the relationship between the total IDC-side loss and the power supply to the IDC, and the regression variance and error shown in Tab. \ref{table.case123_sample} proves the accuracy of the regressed function in this case (the definition of the error here is the deviation between the regressed value and the actual value in the percentage form).} Since the upper layer is an MINLP problem, there is no method can guarantee a global optimal solution as the benchmark. To compare the performance, the time consumption and objective value are shown in Tab. \ref{table.case123_solution_accuracy}, which indicates that the proposed method achieves the better solution with higher efficiency than the conventional one.

\begin{table}[ht]
\centering
\caption{Lower Layer Sampling and Regression Results}
\label{table.case123_sample}
\begin{tabular}{ccccc}
\toprule
 &
  \multicolumn{2}{c}{Sample Points Repredict} &
  \multicolumn{2}{c}{Out-of-Sample Points Predict} \\
\midrule
  \begin{tabular}[c]{@{}c@{}}Sample\\ Number\end{tabular} &
  \begin{tabular}[c]{@{}c@{}}Average\\ Variance\end{tabular} &
  \begin{tabular}[c]{@{}c@{}}Average\\ Error\end{tabular} &
  \begin{tabular}[c]{@{}c@{}}Average\\ Variance\end{tabular} &
  \begin{tabular}[c]{@{}c@{}}Average\\ Error\end{tabular} \\
\midrule
1000 &
  1.07 &
  0.56\% &
  1.09 &
  0.60\%\\
\bottomrule
\end{tabular}
\end{table}

\begin{table}[ht]
\centering
\caption{Performance of Proposed and Conventional Solution Method} 
\label{table.case123_solution_accuracy} 
\begin{tabular}{ccccc}
\toprule
 &
  \multicolumn{2}{c}{Proposed Method} &
  \multicolumn{2}{c}{Conventional Method} \\
\midrule
Solver &
  \begin{tabular}[c]{@{}c@{}}Consumed\\ Time (s)\end{tabular} &
  \begin{tabular}[c]{@{}c@{}}Objective\\ Value\end{tabular} &
  \begin{tabular}[c]{@{}c@{}}Consumed\\ Time (s)\end{tabular} &
  \begin{tabular}[c]{@{}c@{}}Objective\\ Value\end{tabular} \\
\midrule
BARON &
  133 &
  2479k &
  483 &
  2688k\\
\bottomrule
\end{tabular}
\end{table}

\begin{figure}[ht]
\centering 
\includegraphics[width=0.48\textwidth]{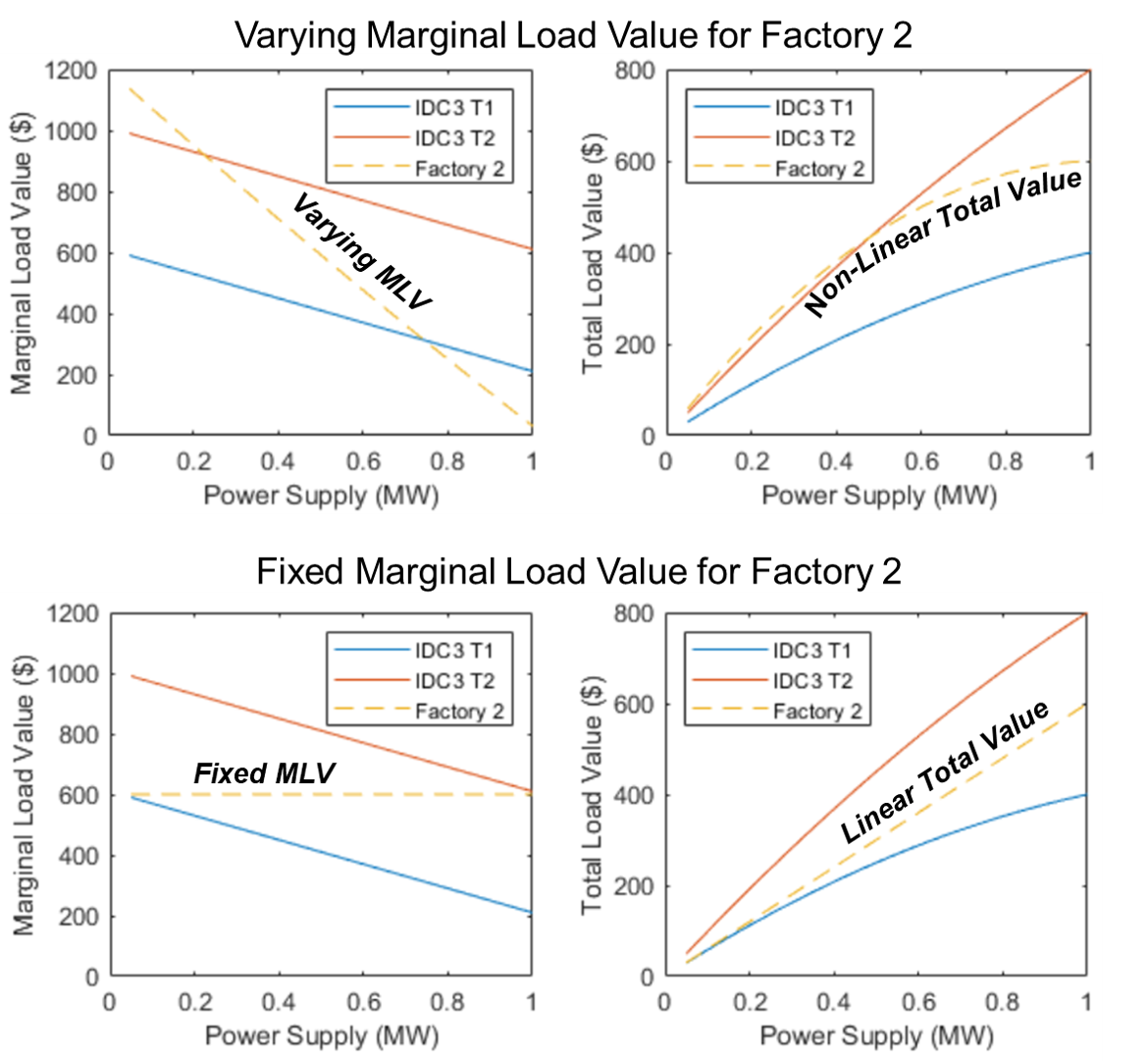}
\caption{Load Value Changing with Power Supply Amount in IDC 3 and Factory 2} 
\label{case123_utility} 
\end{figure}

\noindent \textit{\textbf{Discussion: Considering Varying Marginal Load Value in the Restoration Model}}\\
\hspace*{0.5cm}
In Tab. \ref{table.case123_result}, restoration results show that, with the same line repairing sequence, the proposed model considering varying MLV achieves better solution than the model with fixed MLV. The reason is that varying MLV will change the generation dispatch scheme at different time slots compared to the conventional fixed MLV. For example, before Line 4 is repaired, DG2's generation should be dispatched to restore IDC3 and Factory2. If the varying MLV of non-flexible loads is not considered, the MLV of Factory 2 will be a constant, which is smaller than the MLV of IDC3 at T1 and larger than that at T2 as shown in the lower left diagram of Fig. \ref{case123_utility}. So the dispatch scheme based on fixed MLV shown in the second row of Tab. \ref{table.case123_utility} delivers all power to Factory2 at T1 and to IDC3 at T2. If the varying MLV is considered, the MLV of Factory2 will not be always smaller (T2) or larger (T1) than IDC3 as shown in the upper left of Fig. \ref{case123_utility}, and a more rational dispatch scheme shown in the third row of Tab. \ref{table.case123_utility} will be used to partially restore both IDC3 and Factory2, resulting in a smaller loss than the fixed MLV.

\begin{table}[]
\centering
\caption{Power Supply Allocation Plan and Related Loss of IDC 3 and Factory 2 at Period T1 and T2}
\label{table.case123_utility}
\begin{tabular}{p{0.07\textwidth} p{0.04\textwidth}<{\centering} p{0.04\textwidth}<{\centering} p{0.05\textwidth}<{\centering} p{0.04\textwidth}<{\centering} p{0.04\textwidth}<{\centering} p{0.04\textwidth}<{\centering}}
\toprule
Method &
  Time &
  \begin{tabular}[c]{@{}c@{}}IDC 3\\ Power\\ Supply\\ (MW)\end{tabular} &
  \begin{tabular}[c]{@{}c@{}}Factory 2\\ Power\\ Supply\\ (MW)\end{tabular} &
  \begin{tabular}[c]{@{}c@{}}IDC 3\\ Loss\\ (\$)\end{tabular} &
  \begin{tabular}[c]{@{}c@{}}Factory 2\\ Loss\\ (\$)\end{tabular} &
  \begin{tabular}[c]{@{}c@{}}Total\\ Loss\\ (\$)\end{tabular} \\
\midrule
\begin{tabular}[c]{@{}c@{}}Fixed MLV\\ for Factory 2\end{tabular} &
  \begin{tabular}[c]{@{}c@{}}T1\\ T2\end{tabular} &
  \begin{tabular}[c]{@{}c@{}}0\\ 2\end{tabular} &
  \begin{tabular}[c]{@{}c@{}}1\\ 0\end{tabular} &
  \begin{tabular}[c]{@{}c@{}}400k\\ 0\end{tabular} &
  \begin{tabular}[c]{@{}c@{}}0\\ 600k\end{tabular} &
  1000k \\
\midrule
\begin{tabular}[c]{@{}c@{}}Varying MLV\\ for Factory 2\end{tabular} &
  \begin{tabular}[c]{@{}c@{}}T1\\ T2\end{tabular} &
  \begin{tabular}[c]{@{}c@{}}0.31\\ 0.73\end{tabular} &
  \begin{tabular}[c]{@{}c@{}}0.69\\ 0.27\end{tabular} &
  \begin{tabular}[c]{@{}c@{}}235k\\ 172k\end{tabular} &
  \begin{tabular}[c]{@{}c@{}}62k\\ 321k\end{tabular} &
  790k\\
\bottomrule  
\end{tabular}
\end{table}

\vspace*{0.5\baselineskip} 
\noindent \textit{\textbf{\textcolor{black}{Discussion: Computational Efficiency of the Proposed Solution Method}}}\\
\hspace*{0.5cm}

\textcolor{black}{The computational efficiency of the proposed GPR based solution method is discussed her. Compared to the conventional solution method shown in Eq. \eqref{eq28} - \eqref{eq29}, the proposed solution method shown in Eq. \eqref{eq26} - \eqref{eq27} uses a regressed function $\hat{h}(\cdot)$ to represent the lower layer, which reduces the size of the whole problem and hence increases the computational efficiency. However, the function $\hat{h}(\cdot)$ regressed by GPR is usually nonlinear, which increase the nonlinearity of the problem and therefore slow down the computation. The overall efficiency of the proposed method is determined by the two factors together.} 

\textcolor{black}{In the first case study, because the varying MLV and the network reconfiguration is not considered, the upper layer model is a LP problem with 260 variables, and the lower layer model is a MILP problem with 201 variables as shown in Tab. \ref{table.case33_efficiency}. The size of the whole problem is not large, so the conventional method can solve it within 0.1s. Although the proposed solution method reduces the problem size, but the introduced nonlinearity makes the problem become a NLP problem, which takes 0.7 s to solve. So the computational efficiency of the proposed solution method is not better than the conventional one on simple cases. 
}

\textcolor{black}{To further test the performance of the proposed solution method on large-scaled problems, another simulation is designed based on the second case study. The number of computing workloads in the two IDCs in the second case study is increased from 240, which means a ten-fold increase of the size of the lower layer problem. Now the restoration problem in the extended second case study includes 71,078 variables, and the majority of the variables are from the lower layer model. In the second case study, because the varying MLV and the network reconfiguration are considered, the power system restoration model in the upper layer is a MINLP problem, as well as the integrated model combining the two layers. Obviously, this large-scale MINLP problem is very challenging for the conventional solution method. As shown in Tab. \ref{table.case123_solution_efficiency}, the conventional method does not find a feasible solution within 10,000 s. On the other side, the proposed solution method uses a 12-dimension function to represent the lower layer, which reduces 90\% of the total variables and 98\% of the binary variables. Even though the regressed function increases the nonlinearity of the problem, the proposed method finds a local optimal solution at the 177th second. Considering the large-scale MINLP is too difficult to solve, the varying MLV is linearized for the conventional method, so the original MINLP problem is simplified as a MILP problem with the same number of variables. The conventional method spends 34 seconds to find the optimal solution of the linearized problem. However, because of the error brought by the linearization is too large, the obtained solution is worse than the proposed solution method. Based on this, a conclusion can be drawn that the proposed GPR based solution method is more efficient than the conventional method when dealing with the large-scale problems.
}

\begin{table}[ht]
\centering
\caption{\textcolor{black}{Performance of Proposed and Conventional Solution Method}}\label{table.case123_solution_efficiency}
\begin{tabular}{c|ccccc}
\toprule
\multicolumn{2}{c}{Method} &
  \begin{tabular}[c]{@{}c@{}}Total\\ Variable\\ Number\end{tabular} &
  \begin{tabular}[c]{@{}c@{}}Binary\\ Variable\\ Number\end{tabular} &
  \begin{tabular}[c]{@{}c@{}}Consumed\\ Time (s)\end{tabular} &
  \begin{tabular}[c]{@{}c@{}}Total\\ Loss (\$)\end{tabular} \\
\midrule
\multirow{3}{*}{\begin{tabular}[c]{@{}c@{}}Proposed\\ Method\end{tabular}} &
  Upper &
  6856 &
  976 &
  \multirow{3}{*}{177} &
  \multirow{3}{*}{2,669k} \\
 &
  Lower &
  12 &
  0 &
   &
   \\
 &
  Total &
  6868 &
  976 &
   &
   \\
\midrule
\multirow{3}{*}{\begin{tabular}[c]{@{}c@{}}Conventional\\ Method\end{tabular}} &
  Upper &
  6856 &
  976 &
  \multirow{3}{*}{\textgreater{}10000} &
  \multirow{3}{*}{\begin{tabular}[c]{@{}c@{}}No\\ Solution\\ Found\end{tabular}} \\
 &
  Lower &
  67,212 &
  38,400 &
   &
   \\
 &
  Total &
  74,068 &
  39,376 &
   &
   \\
\midrule
\multirow{3}{*}{\begin{tabular}[c]{@{}c@{}}Linearized\\ Convetional\\ Method\end{tabular}} &
  Upper &
  6856 &
  976 &
  \multirow{3}{*}{34} &
  \multirow{3}{*}{2,952k} \\
 &
  Lower &
  67,212 &
  38,400 &
   &
   \\
 &
  Total &
  74,068 &
  39,376 &
   &\\
\bottomrule
\end{tabular}
\end{table}


\section{Conclusions}
Bilevel optimization problems widely arise in the power system, but most existing approaches rely on the assumption of the leader's omniscience, which is untenable in many cases. This paper proposes a new method for bilevel optimization in the power system to fill this gap, and introduces an advanced bilevel model for post-event CL restoration as a testbed. The proposed method uses Gaussian Process Regression to approximate how the followers respond to the operation decisions of the power system. This approximated function then is provided to the power system to make decisions without requiring details of the followers. On the other hand, the advanced bilevel CL restoration model takes the load-side operation and the varying marginal load value into consideration, which can increase the optimality of the restoration solutions but brings more challenges in solving the model. The simulation results of two case studies not only show the effectiveness of the CL restoration model, but also prove that the proposed solution method can solve bilevel problems more efficiently than conventional methods, and the advantages become more obvious when the problem complexity increases. In sum, the proposed solution method provides a new direction for solving various collaborative bilevel optimization problems in the power system, and owns a wider application range and higher computational efficiency than existing methods.

\bibliographystyle{IEEEtran}
\bibliography{main.bbl}

\end{document}